\pdfoutput=1
\documentclass[11pt]{article}
\usepackage{EMNLP2022}

\usepackage{times}
\usepackage{latexsym}
\usepackage[T1]{fontenc}
\usepackage[utf8]{inputenc}
\usepackage{microtype}

\usepackage{graphicx}
\usepackage{tabularx}
\usepackage{amsmath}
\usepackage{amsthm}
\usepackage{amsfonts}
\usepackage{multirow}
\usepackage{booktabs}
\usepackage{color}
\usepackage{amssymb}
\usepackage{float}

\title{Understanding Social Media Cross-Modality Discourse in Linguistic Space}

\author{Chunpu Xu\textsuperscript{\rm 1}, Hanzhuo Tan\textsuperscript{\rm 1}, Jing Li\textsuperscript{\rm 1}\thanks{~~~Corresponding author}, 
Piji Li\textsuperscript{\rm 2}\\ 
\textsuperscript{\rm 1} Department of Computing, The Hong Kong Polytechnic University\\
\textsuperscript{\rm 2} College of Computer Science and Technology, \\
Nanjing University of Aeronautics and Astronautics\\
\textsuperscript{\rm 1}\texttt{\{chun-pu.xu,han-zhuo.tan\}@connect.polyu.hk}\\ \textsuperscript{\rm 1}\texttt{jing-amelia.li@polyu.edu.hk;}\textsuperscript{\rm 2}\texttt{pjli@nuaa.edu.cn}}

\begin{document}
\maketitle
\begin{abstract}
The multimedia communications with texts and images are popular on social media.
However, limited studies concern how images are structured with texts to form coherent meanings in human cognition.
To fill in the gap, we present a novel concept of cross-modality discourse, reflecting how human readers couple image and text understandings.
Text descriptions are first derived from images (named as \textbf{subtitles}) in the multimedia contexts. 
Five labels -- entity-level \textit{insertion}, \textit{projection}  and \textit{concretization} and scene-level \textit{restatement} and \textit{extension} --- are further employed to shape the structure of subtitles and texts and present their joint meanings.
As a pilot study, we also build the very first dataset containing 16K multimedia tweets with manually annotated discourse labels.
The experimental results show that the multimedia encoder based on multi-head attention with captions is able to obtain the-state-of-the-art results.
\end{abstract}

\section{Introduction}
\begin{figure*}[t]
\centering
\includegraphics[width=0.9\textwidth]{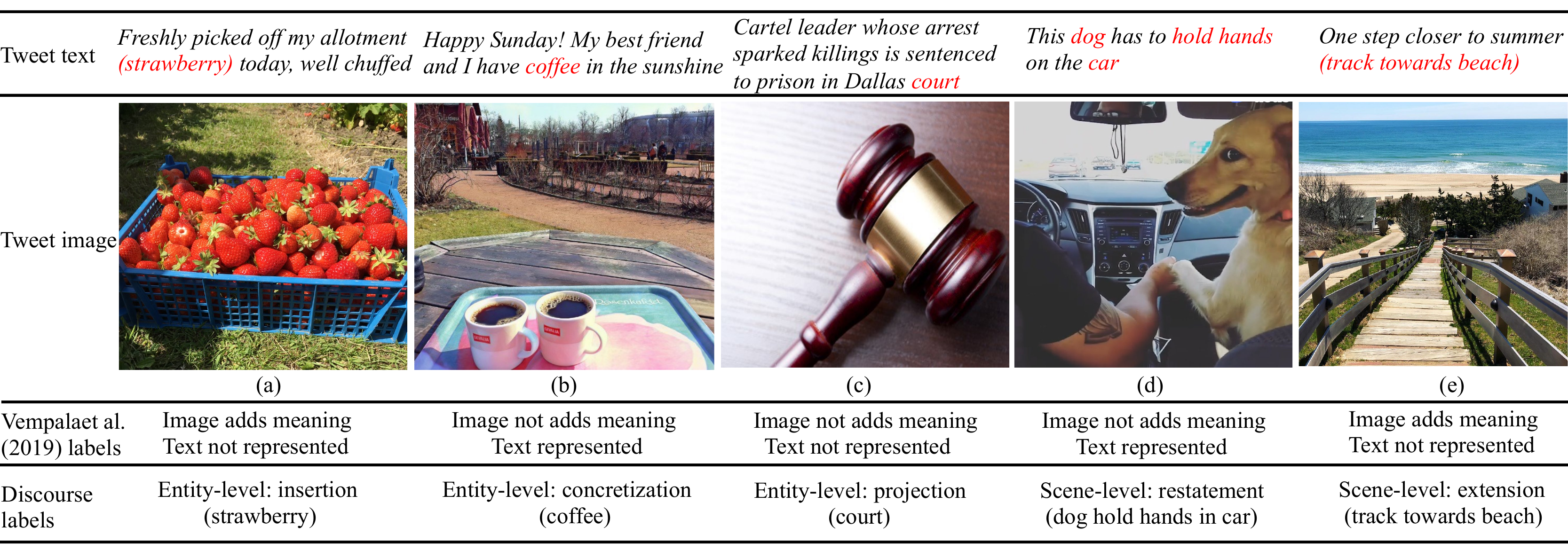}
\vspace{-1em}
\caption{The five cross-modality discourse labels and their examples. The rows from top to bottom display their texts, images, the image-text relation labels in \citet{DBLP:conf/acl/VempalaP19}, and our cross-modality discourse categories. 
The labels in \citet{DBLP:conf/acl/VempalaP19} concern whether new meanings are added by images to texts, whereas
ours define the linguistic roles of images and their pragmatic relations with texts for coherence.}
\vspace{-1em}
\label{fig:case}
\end{figure*}

The growing popularity of multimedia is revolutionizing the  communications on social media.
The conventional text-only form has been expanded to cross modalities involving texts and images in information exchange. 
For multimedia messages, the language understanding acquires more than making sense of both visual and textual semantics; it also matters to figure out what glues them together to exhibit the coherent senses in human's mind.

Nevertheless, most progress made in social media language understanding relies on texts to learn the message-level semantics~\cite{DBLP:conf/emnlp/ShenZHC18,DBLP:conf/emnlp/NguyenVN20}, largely ignoring the rich meanings conveyed in images \cite{DBLP:conf/acl/CaiCW19,DBLP:conf/emnlp/WangLLK20}.
Other recent multimodal studies focus on model designs to combine visual and textual signals  \cite{DBLP:journals/pami/ParkKK19,DBLP:conf/acl/LiZZWLJC20, yu-etal-2021-vision}, ignoring the insights from how humans understand the implicit structure underlying a multimedia post.

In light of these concerns, we consider images as an integral part of social media language and propose a novel concept of \textit{cross-modality discourse}, which defines how human readers structure the coherent meanings from image and text modalities. 
Our work is inspired by \citet{DBLP:conf/acl/VempalaP19} examining the information overlap between images and texts, whereas we take a step further to characterize how multimedia messages make sense to humans, which is beyond a simple yes-or-no prediction to whether new thing is observed.
To the best of our knowledge, \emph{we are the first to extend discourse --- a pure linguistic concept --- to define the linguistic roles played by images and their pragmatic relations with texts to shape the coherent meanings.
} 

In general, cross-modality discourse is defined by the operations adopted in human perception to couple image and text semantics. 
Readers may first extract the information from the images acquired to complete the cross-modality understanding, either in form of the local objects (entities) or global scenes \cite{rayner200935th}.
Then, the extracted entities or scenes are represented in texts, named as the images' \textbf{subtitles}, which can further contribute to structure the entity-level or scene-level discourse with the matching texts in the multimedia contexts.
Concretely, for entity-level discourse, it is detailed into \emph{insertion}, \emph{projection}, and \emph{concretization}, 
according to whether the entity is omitted, described, or mapped; 
similarly, scene-level \emph{restatement} and \emph{extension} are employed to reflect whether the story in one modality recurs or continues in the other.

To illustrate the definitions above, Figure \ref{fig:case} shows five multimedia Twitter posts. 
As can be seen from (a), readers may concentrate on the object ``strawberry'' and \emph{insert} its name into the texts omitting the entity.
As for (b), the ``coffee'' object should be extracted from the image to \emph{concretize} the word ``coffee'' in the text.
In (c), the word ``court'' in text is linked with the ``gavel'' object.
The image in (d) helps \emph{restate} the texts scene (a dog holds hands in the car).
In (e), the global scene works as an \emph{extension} to texts and completes the story: ``We are one step closer to summer following the track towards beach.''.   

On the contrary, the image-text relations in \citet{DBLP:conf/acl/VempalaP19} are limited to whether images add new meanings to texts, which is nonetheless insufficient to reflect how language is understood in multimedia contexts.

\emph{As a pilot study of cross-modality discourse, we also present the very first dataset to explore the task.}
It is collected from Twitter and contains  16K high-quality multimedia posts with manual annotations on their discourse labels.\footnote{The dataset and code are released at \url{https://github.com/cpaaax/Multimodal_Discourse}.}
We believe our task and the associated dataset, being the first of its kind, will be potentially beneficial to help machines gain the ability to understand social media language with multimodal elements. 

To that end, we present a framework to learn the discourse structure across texts and images.
Inspired by the recent advances in multimodal learning~\cite{DBLP:conf/emnlp/WangLLK20, yu2020improving}, we employ the multi-head attention mechanism \cite{DBLP:conf/nips/VaswaniSPUJGKP17} to explore the visual-textual  representations reflecting cross-modality interactions. 
Besides, to characterize subtitles for discourse learning, image captions generated from model trained on COCO captioning dataset~\cite{DBLP:conf/eccv/LinMBHPRDZ14} are leveraged as additional features.

For empirical studies on cross-modality discourse, we conduct comprehensive experiments on our dataset.
The comparison results on classification show the challenges for machines to infer discourse structure and it is beyond the capability of advanced multimodal encoders 
to well handle our task.
Nevertheless, exploring correlations of texts, captions, and visual-textual interactions helps exhibit the state-of-the-art performance in both the intra-class and overall evaluation.
We further examine the effects of varying modalities and text length and find that text signals are crucial for discourse inference while joint effects of texts, images, and captions present the best results.
At last, the qualitative analysis demonstrates how the multi-head attention in our model interprets discourse structure.

\section{Related Work}

Our paper crosses the lines of multimedia learning and discourse analysis in natural language processing. Here comes more details.

\vspace{-0.5em}

\paragraph{Multimedia Learning.} Our paper is in the line with cross-media research that attempts to fuse textual and visual features. 
There are various deep learning methods proposed to leverage crossmodal features, either based on advanced neural architectures like co-attentions~\cite{DBLP:conf/eccv/XuS16,DBLP:conf/nips/LuYBP16} and multi-head attentions~\cite{DBLP:conf/nips/VaswaniSPUJGKP17,DBLP:conf/emnlp/WangJLKXH20}, or pre-trained visual-lingual representations~\cite{DBLP:conf/nips/LuBPL19,DBLP:conf/iclr/SuZCLLWD20, DBLP:conf/cvpr/ZhangLHY0WCG21}.
Their effectiveness are demonstrated in both conventional vision-language tasks, such as image captioning \cite{DBLP:journals/pami/ParkKK19,DBLP:conf/aaai/ZhouPZHCG20,shi-etal-2021-enhancing} and visual question answering (VQA)~\cite{DBLP:conf/cvpr/Yu0CT019,DBLP:conf/emnlp/TanB19, si-etal-2021-check}, and  social media applications, such as sarcasm detection~\cite{DBLP:conf/acl/CaiCW19}, event tracking~\cite{DBLP:conf/acl/LiZZWLJC20,DBLP:conf/cvpr/AbavisaniWHTJ20}, keyphrase prediction~\cite{DBLP:conf/aaai/Zhang00TY019,DBLP:conf/emnlp/WangLLK20}.

It is seen that most progress to date made in this line focus on advancing methodology designs for general purposes~\cite{DBLP:conf/iclr/SuZCLLWD20,DBLP:conf/aaai/ZhouPZHCG20} or specific applications~\cite{DBLP:conf/emnlp/WangLLK20} to better capture the matched semantics across varying modalities.
However, their effectiveness over social media data would be inevitably compromised resulted from the intricate image-text interactions~\cite{DBLP:conf/acl/VempalaP19}.
We thus borrow the insights from human perception to interpret image-text relations from the linguistic viewpoints and propose the task to learn discourse structure in multimedia contexts.
It is a fundamental research exhibiting the potential to help the models gather cross-modality understanding capability and might benefit various downstream applications.

We are also related with previous categorization tasks on social media to understand image-text relations, such as information overlap \cite{DBLP:conf/acl/VempalaP19}, point-of-interest types \cite{DBLP:conf/emnlp/VillegasA21}, author purposes \cite{DBLP:conf/emnlp/KrukLSLJD19}, object possessions \cite{DBLP:conf/emnlp/ChinnappaMB19}, and so forth.
Besides, interestingly, the ``discourse'' concept is also employed to examine the image-text relations in cooking recipes~\cite{DBLP:conf/naacl/AlikhaniCMS19}.
Compared with these studies concatenating visual and textual embeddings in a ``common'' space, we craft text-formed subtitles to convey visual stories and explore how they shape the coherent meanings with the post texts in linguistic space.
It will consequently allow deep semantic learning to capture the implicit structure holding image and text modalities, while the existing models might be incapable to gather senses of language understanding via simple feature concatenation.

\vspace{-0.5em}
\paragraph{Discourse Analysis.} This work is related to prior studies on text-level discourse structures.
The popular tasks in the styles of either RST (Rhetorical Structure Theory) \cite{mann1988rhetorical,DBLP:conf/emnlp/LiuLJHB19} or PDTB (Penn Discourse Tree Bank) \cite{DBLP:conf/lrec/PrasadDLMRJW08,DBLP:conf/emnlp/XuHRYZZ18} explore the rhetorical relations of discourse units (e.g., phrases or sentences) that cohesively connect them form a sense of coherence.  
These studies have demonstrated their helpfulness in diverse stream of NLP applications \cite{DBLP:conf/acl/ChoubeyLHW20}, such as sentiment analysis \cite{DBLP:conf/emnlp/BhatiaJE15}, text categorization~\cite{DBLP:conf/acl/JiS17}, and microblog summarization~\cite{DBLP:journals/coling/LiSWW18}. 
Nevertheless, limited work examines a social media image as a discourse unit of the pragmatic structure in multimedia contexts, which is a gap to be filled in this work.
\section{Study Design}

In this section, we first define the task to predict cross-modality discourse in $\S$\ref{ssec:preliminary:task}. Then, we introduce how we construct the dataset in $\S$\ref{ssec:preliminary:collection}, followed by the data analysis in $\S$\ref{ssec:preliminary:analysis} and the potential applications in  $\S$\ref{ssec:study-application}.

\subsection{Task Definition}\label{ssec:preliminary:task}
In our task, the input is an image-text pair from a multimedia post on social media, following the previous practice~\cite{DBLP:conf/acl/VempalaP19}. 
For each pair, the goal is to output a label from a predefined set that 
cover the major categories of cross-modality discourse on social media. 
Our intuition is that images are relatively more eye-catching and likely to be processed before the texts.
For image understandings, the previous findings from psychological experiments~\cite{rayner200935th} point out that humans may first recognize and extract the meanings from global scenes to fill the information gap in contexts; if the gap still exists, they may go back to capture the local objects.
Based on that, we first coarsely categorize the discourse label set into the level of entity (object) and scene, depending on whether object or scene is extracted to make sense of the joint meanings of images and texts.

To further elaborate the label design, the extracted information from an image (as an object or scene) is mapped to the text modality to form the \textbf{subtitle}, which allows us to formulate how human senses structure the coherent meaning with subtitles and post texts.

For entity-level discourse, three cases are examined: the entity is omitted, mentioned or linked in the texts.
For the absent entity (e.g., Fig. \ref{fig:case}(a)), the subtitle, in form of entity name, should be \emph{inserted} into the post text to complete the  meanings of a message, while the entity in Fig. \ref{fig:case}(b) is \emph{concertized} by the object in images.
And the entity in Fig. \ref{fig:case}(c) is implicitly
projected into the relevant object.
We henceforth design entity-level \textbf{insertion},    \textbf{concretization}, and \textbf{projection} to describe the above three cases, respectively.
%


Similarly, scene-level discourse can be separated into \textbf{restatement} and \textbf{extension} categorizes.
The former refers to image serving as texts description (e.g., Fig. \ref{fig:case}(d)) and for the latter, posts presenting image scenes to elaborate the story left as white space in the texts (e.g., Fig. \ref{fig:case}(e)).

\subsection{Data Collection and Annotation}\label{ssec:preliminary:collection}

\begin{figure}[t]
\centering
\includegraphics[width=1\columnwidth]{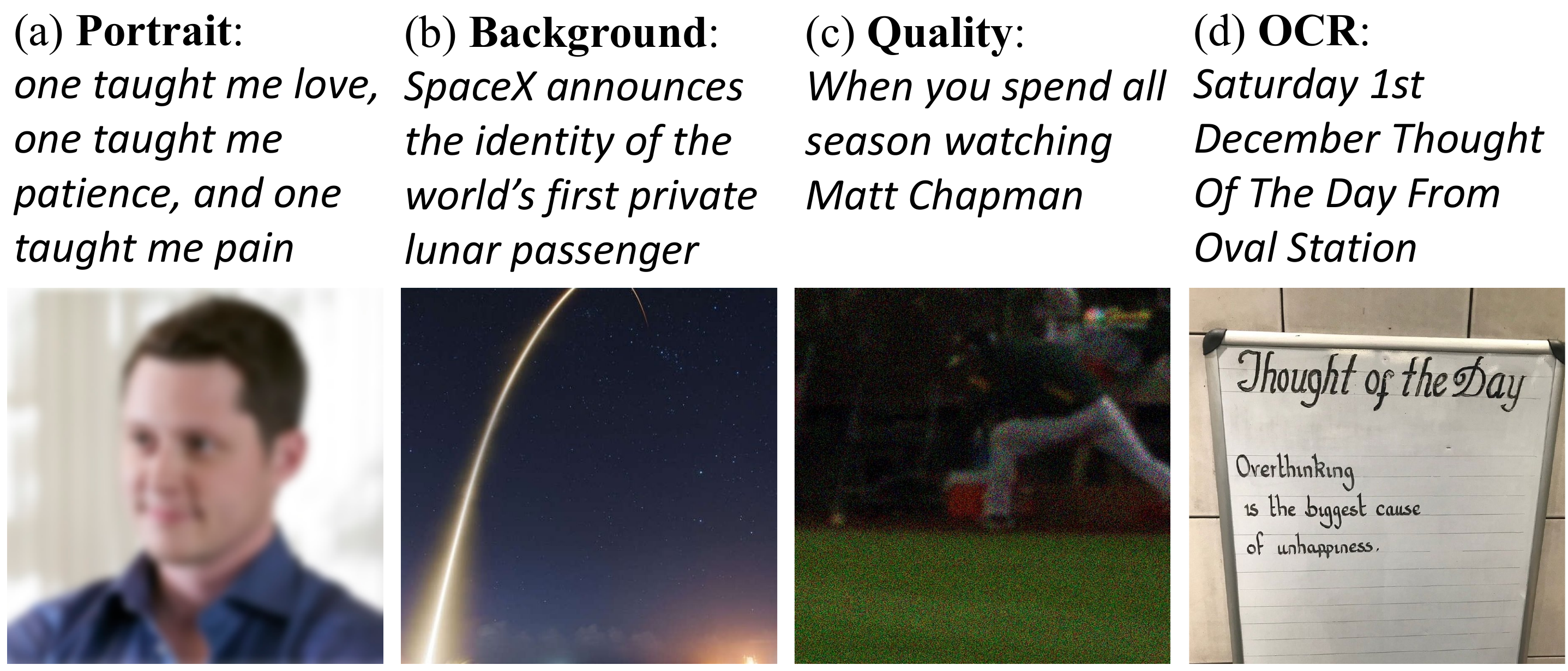}
\vspace{-2em}
\caption{Examples tweets of the four ``bad'' types.
(a) \textbf{Portrait} image with quotes in texts.
(b) \textbf{Background} is externally required for understanding (rocket trajectory scenes here). 
(c) \textbf{Low-quality} image where objects could be barely observed. 
(d) \textbf{OCR Subtitle} (``Thought Of The Day") appear in the image in optical characters.
}
\label{fig:dataBadCase}
\end{figure}

Our dataset is gathered from Twitter\footnote{\url{twitter.com}}, which is drawing attentions to research digital communications \cite{mozafari2019bert,nikolov2019nikolov,muller2020covid} and exhibits prominent use of multimedia posts \cite{DBLP:conf/acl/VempalaP19,DBLP:conf/emnlp/WangLLK20}. 
We first crawled the raw data using Twitter streaming API\footnote{\url{developer.twitter.com/en/docs/tutorials/stream-tweets-in-real-time}} and removed non-English posts and those with texts only or multiple images.
Afterwards, to better model discourse from the noisy Twitter data \cite{DBLP:conf/acl/VempalaP19}, we 
removed samples that might hinder the learning of non-trivial discourse signals.
Here, four types of ``bad'' image-text pairs might provide tremendous noise in the learning, which are shown in Fig. \ref{fig:dataBadCase}.

The first type refers to image portraits with some quotes to share the insights of life (henceforth \textbf{portraits}), where images and texts are not coherently related (from linguistic viewpoints) and discourse structure are unable to be defined for them.
Moreover, many of them contain authors' selfies, which might raise privacy concerns.
The second type of posts, namely \textbf{background}, relies on external knowledge to capture the meanings (e.g., Fig. \ref{fig:dataBadCase}(b)), which is beyond the capability of language understanding given only the images and the matching texts. 
For the third, we consider \textbf{low-quality images} (e.g., low resolution and blurred ones like Fig. \ref{fig:dataBadCase}(c)), from which it is hard to capture the visual meanings.
The last one refers to \textbf{OCR subtitles} (Fig. \ref{fig:dataBadCase}(d)), where the subtitles appear in the images as optical characters. It may result in a degeneration of cross-modality discourse to text-level discourse and render the learning of trivial features.

In the data annotation, we first selected 25 typical examples corresponding to each discourse label and provide them together with the annotation guidelines (with the detailed description of each label) for quality control. 
Then, two postgraduate students majoring in linguistics were recruited to manually label the discourse categories, given an image-text pair. 
``Bad'' samples falling in the above four types should also be indicated in the annotation process. 
The inter-annotator agreement is 79.8\% and we only kept the data with labels agreed by both annotators to ensure the feature learning quality in noisy data.
At last, posts in ``bad'' types were taken away and the final dataset presents 16k multimedia tweets with manual labels in five discourse categories.

\begin{table}[t]
	 \centering
	 
{\renewcommand{\arraystretch}{0.1}
\resizebox{0.9\columnwidth}{!}
{
	\begin{tabular}[b]{l||c||c|c|c|c|c}
		\toprule
		 &\textbf{Total}  &\textbf{Ins} &\textbf{Con} &\textbf{Pro} &\textbf{Res} &\textbf{Ext}  \\
		\midrule
		Num  
		& 16,000  & 839  & 10,558  & 690 & 1,826 &2,087\\
		 Len
		& 10.69 & 9.11   & 10.85  & 10.98 & 11.24 &9.92 \\
		\bottomrule	\end{tabular}}}
	\vspace{-0.5em}
	\caption{Statistics of the total data and that with each label: Ins: Insertion; Con: Concretization; Pro: Projection; Res: Restatement; Ext: Extension. Len: average word number in texts. Num: tweet number.
	}

	\label{tab:stat}
	\vspace{-1em}
\end{table}

\subsection{Data Analysis}\label{ssec:preliminary:analysis}
Here we conduct a preliminary analysis of our dataset and show the statistics in Table \ref{tab:stat}. 
There exhibits imbalanced labels, where concretization and extenstion labels are relatively more popular compared to the other three.
This indicates the diverse preferences of Twitter users in the way they choose to  structure texts and images and the potential challenge for models to handle our task.

For the text length, it is seen that most tweets contain limited words, challenging the models to capture essential features from textual signals. 
Interestingly, we compare our statistics with other text-only Twitter datasets in previous work ~\cite{DBLP:conf/naacl/WangLKLS19} and find our multimedia tweets have 30\% fewer words on average. 
This implies that authors may tend to put less content in the text of multimedia posts, and figure the missing information in images for compensation.
We also notice that insertion and extension discourse exhibit relatively shorter texts on average.
It is probably because they exhibit the omitted content in texts, which presents in images.

\begin{figure}[t]
\centering
\includegraphics[width=0.9\columnwidth]{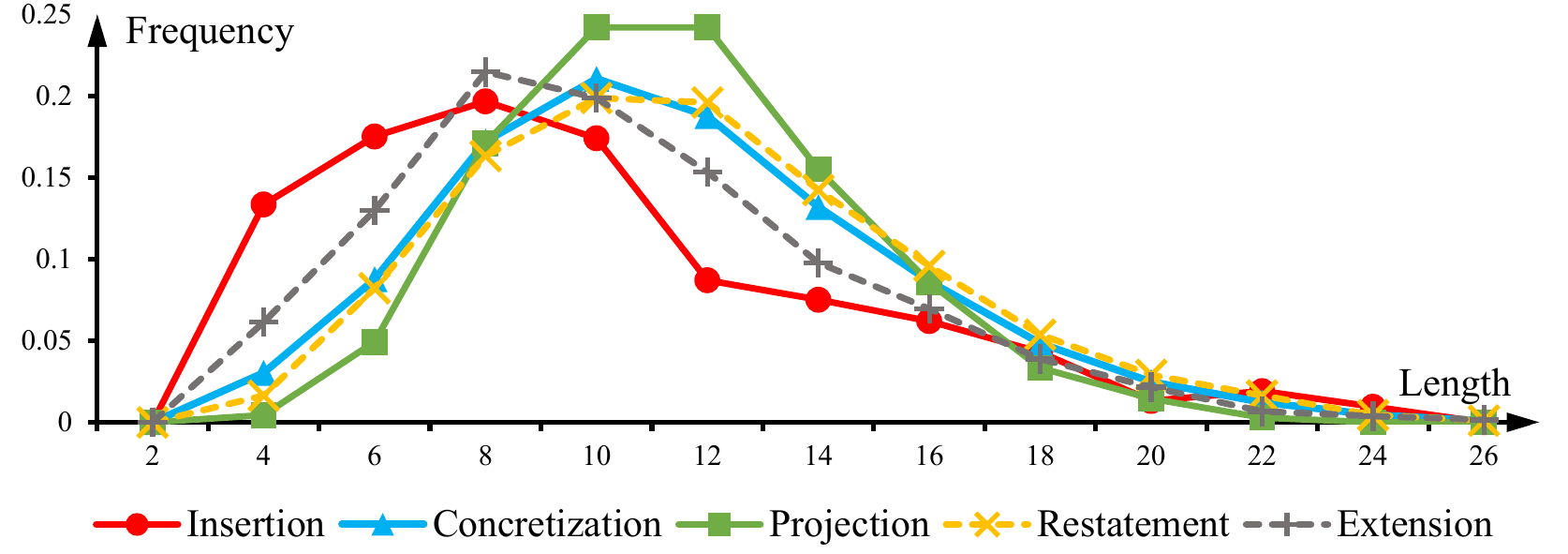}
\vspace{-1em}
\caption{
Text lengths (token number) distribution of
posts with varying discourse labels.
}
\vspace{-1em}
\label{fig:len}
\end{figure}

To further characterize text length in our dataset, Fig. \ref{fig:len} shows the word number distribution of tweet texts with varying labels. 
All the curves demonstrate the sparse distribution over text length, owing to the freestyles of social media writings.
Insertion and extension curves first peak at 8 words while the others at 10-12, then all present long tails afterwards. 
This again shows that texts in multimedia posts may provide limited content and those in insertion and extension contain fewer words.

\subsection{Potential Applications}\label{ssec:study-application}
In this subsection, we further discuss the potential downstream applications of our task and datasets, which might inspire the design of future work.
A straightforward application is microblog summarization --- an important task to distill the salient content from massive social media data.
As many state-of-the-art summarization models only allow textual input while multimedia posts are prominent on social media, it may require the compression of these posts into text for easy processing.
It is different from the traditional image captioning task \cite{anderson2018bottom,rennie2017self,huang2019adaptively}, where the generated captions are translated from images. 
For a social media post, the text cannot trivially be seen as a ``translation'' of image, because of possibly ambiguous image-text interactions therein.
Considering crucial roles played by discourse analysis in summarization \cite{DBLP:conf/acl/XuGCL20}, it is not hard to envision that our cross-modality discourse, describing how image and text structure coherence, would contribute to the research of multimedia summarization.

In addition, cross-modality discourse can be viewed as a fundamental task and might be helpful to other downstream tasks  on social media (e.g., multimodal NER \cite{yu2020improving}, multimodal crisis events classfication \cite{abavisani2020multimodal}, multimodal sarcasm detection \cite{cai2019multi},
multimodal sentiment analysis\cite{truong2019vistanet}, and multimodal hashtag prediction \cite{wang2020cross}). 
However, most previous efforts focus on the leverage of visual and lingual representations yet ignore the linguistic essence that glue the two modalities. 
Recently, some work propose multitask learning to consider image-text relations
in multimodal learning. 
For example, \citet{DBLP:conf/aaai/0006W0SW21} investigate the relation propagation between text and image to improve the accuracy of NER in tweets. \citet{DBLP:conf/emnlp/JuZXLLZZ21} utilize multimodal relation types as auxiliary labels to explore multimodal aspect-sentiment analysis. 
The positive results from these studies imply the potential of cross-modality discourse (as a linguistic description of image-text relations) to benefit a wide range of multimodal applications.
Besides, the training data of image-text relation used in \cite{DBLP:conf/aaai/0006W0SW21, DBLP:conf/emnlp/JuZXLLZZ21} is the TRC dataset proposed by \citet{DBLP:conf/acl/VempalaP19}. 
Compared to the TRC dataset, our proposed discourse dataset exhibits a tremendously larger scale (i.e., 16K VS 4.5K) and fine-grained labels for image-text relation, as shown in Fig. \ref{fig:case}.
We therefore believe our dataset would also helpfully advance the performance of various multimodal models.
\section{The Discourse Learning Framework}

\begin{figure}[t]
\centering
\includegraphics[width=0.9\columnwidth]{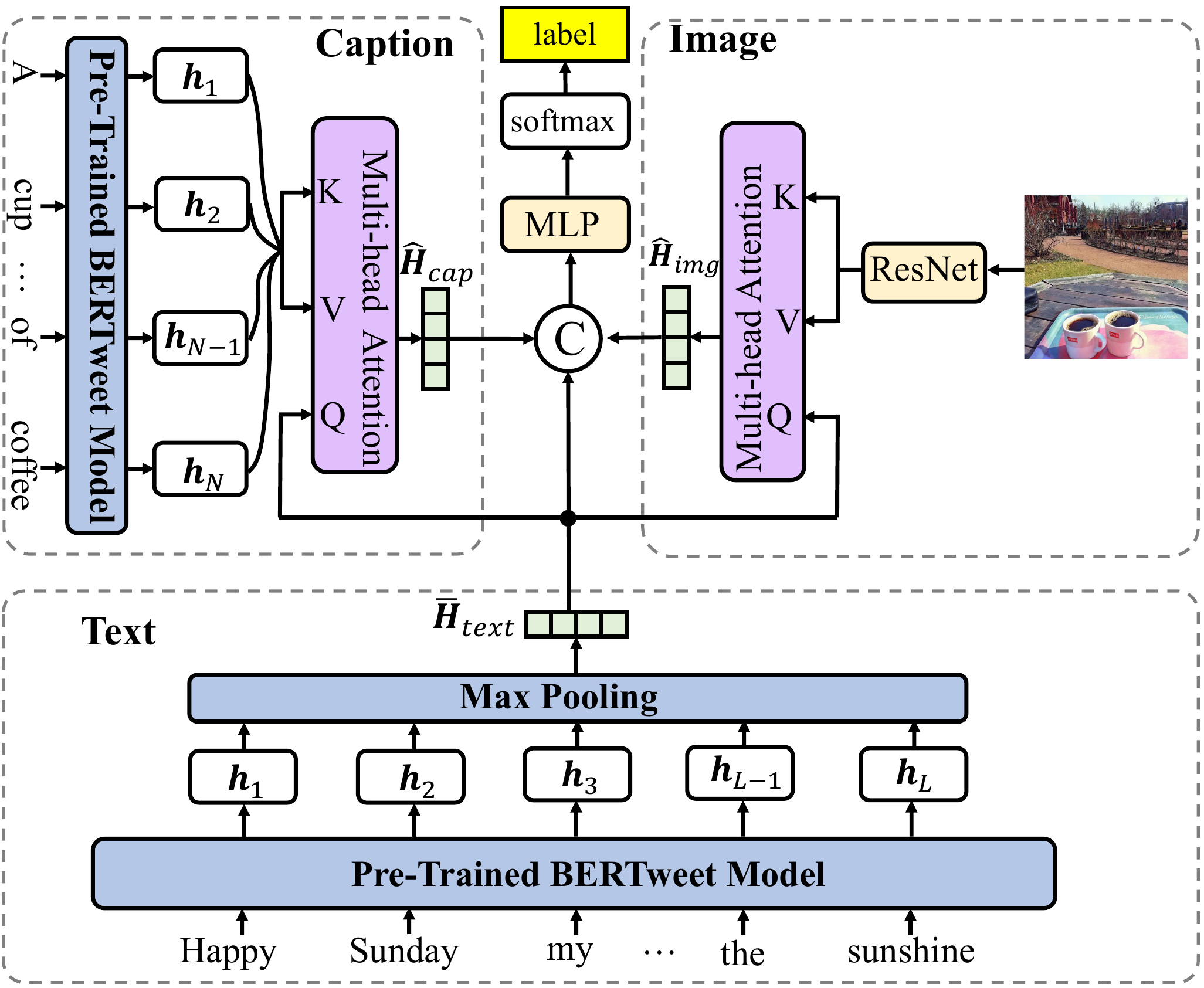}
\vspace{-1em}
\caption{Our framework to learn cross-modality discourse via representations encoded from texts (bottom), captions (upper left), and images (upper right).
The encoded captions and texts are compared at output layer in visual-textual contexts.
}
\label{fig:model}
\vspace{-1.5em}
\end{figure}

In this section, we describe our framework that couples the signals from images and texts to predict their discourse labels. 
As shown in Fig. \ref{fig:model},
the model architecture leverages representations learned from texts, images, and image captions (to reflect subtitles), which will be introduced in $\S$\ref{ssec:model:encoding}. 
Then, we will discuss how we combine multi-modality representations $\S$\ref{ssec:model:integration}.
At last, $\S$\ref{ssec:model:training} presents how we predict the discourse labels and design the training processes.



\subsection{Encoding Text, Image, and Captions}\label{ssec:model:encoding}

\paragraph{Texts Encoding.} 
Here we describe how to learn text features. 
The text encoder is based on the bottom 6-layers of pre-trained Bertweet \cite{DBLP:conf/emnlp/NguyenVN20}.
It is fed with an $L$-length token sequence and
embed its representations into a sequential hidden states $\textbf{H}_{text}=(\textbf{h}_{1},...,\textbf{h}_{L})$, where each element reflects a token embedding.
$\textbf{H}_{text}$ further goes through a max-pooling layer and produces  $\bar{\textbf{H}}_{text}$ to represent the text.

\paragraph{Image Encoding.}
To explore visual signals, images are encoded by CNN-based ResNet-101 \cite{he2016deep} pre-trained on ImageNet \cite{russakovsky2015imagenet}. 
The output of the last convolutional layer in ResNet-101 is extracted as the representation of the input image. 
The size of the feature map is first reduced to $M \times M \times 2048$ and then reshaped into $M^{2}\times2048$. Each $1\times2048$ vector represents the visual features in a corresponding image area and is projected to the same dimension of text feature \textbf{h} by liner layer. The post-level visual feature is denoted as $\textbf{H}_{img}=( \textbf{v}_{1},...,\textbf{v}_{M^{2}} )$, where $\textbf{v}_i$ refers to an $1\times2048$ vector that represents the feature of an area in the image.

\paragraph{Image Caption Encoding}\label{ssec:model:caption}

In order to capture more semantic information from images, we further exploit image captions (henceforth captions) as an additional modality.
Our intuition is that captions may inject essential visual semantics underlying images into a descriptive language in texts \cite{xu2015show}. 
They are potentially helpful to reflect the rich interactions between image objects and discover subtitle-style clues as essential discourse indicators.
We first employ the model presented by \citet{anderson2018bottom} to predict the captions of each image.
The captioning model is pre-trained on the COCO captioning dataset~\cite{DBLP:conf/eccv/LinMBHPRDZ14}, which mostly consists of natural pictures outside social media domain.
Then, we encode the token sequence of captions following the same process of text encoding (discussed above) and yield caption representation: $\textbf{H}_{cap}=( \textbf{h}_{1},...,\textbf{h}_{N})$. Here $N$ indicates the number of tokens in the caption, $\mathbf{h}_i$ refers to the $i$-th hidden state of the Bertweet encoder.

\subsection{Integrating Multimodal Representations}\label{ssec:model:integration}

As pointed out in previous work \cite{DBLP:conf/emnlp/WangLLK20}, modalities on social media data exhibit much more intricate interactions compared with the widely-studied vision-language datasets~\cite{lin2014microsoft,young2014image}. 
To allow the framework to attend various types of cross-modality interactions, we employ multi-head attentions~\cite{DBLP:conf/nips/VaswaniSPUJGKP17} to comprehensively explore the interactions between the encoded image features ($\textbf{H}_{img}$) and max-pooled text representations ($\bar{\textbf{H}}_{text}$).


Concretely, we set text features as the query $\textbf{Q}$, image features as the key and value $\textbf{K}, \textbf{V}$, and compute the multi-head attention $MA(\cdot)$ as follows:
\vspace{-0.5em}
\begin{equation}\small
    MA(\textbf{Q},\textbf{K},\textbf{V})= [hd_{1};...;hd_{n}]\textbf{W}^{O}
\end{equation}

\noindent where $n$ is the number of heads, $[\cdot]$ indicates the concatenation operations, and the attention of the $j$-th head is:
\vspace{-0.5em}
\begin{equation}\small\label{eq:head-attention}
    hd_{j}=A(\textbf{Q}\textbf{W}_{j}^{Q},\textbf{K}\textbf{W}_{j}^{K},\textbf{V}\textbf{W}_{j}^{V})
\end{equation}
\vspace{-1em}
\begin{equation}\small\label{eq:head-attention}
    A(\textbf{Q},\textbf{K},\textbf{V})=\theta(\frac{ \textbf{Q} \textbf{K}^{T} }{ \sqrt{d_k} } )\textbf{V}
\end{equation}


\noindent Here $d_k$ is the normalization factor, $\theta(\cdot)$ means softmax function. $\textbf{W}^{O}, \textbf{W}_{j}^{Q}, \textbf{W}_{j}^{K}, \textbf{W}_{j}^{V}$ are learnable variables. The attended images (in aware of texts) are denoted as $\hat{\textbf{H}}_{img}$, which further serves as the context to help explore the discourse clues from captions and texts.

For discourse modeling, the encoded texts  ($\bar{\textbf{H}}_{text}$) are compared with captions (carrying subtitle-style features) to infer how the subtitles can be structured with texts. 
To that end, we first employ a multi-head attention mechanism to encode text-aware attended caption $\hat{\textbf{H}}_{cap}$, which captures salient contents from captions to indicate discourse categories. 
Furthermore, $\hat{\textbf{H}}_{cap}$ are concatenated with $\bar{\textbf{H}}_{text}$ to model their structure; also concatenated are the attended images $\hat{\textbf{H}}_{img}$ as the image-text interaction contexts for cross-modality discourse learning.




\subsection{Discourse Prediction and  Model Training}\label{ssec:model:training}

The discourse labels are predicted with a multi-layer perceptron (MLP) fed with   $\textbf{H}=[\hat{\textbf{H}}_{cap};\bar{\textbf{H}}_{text};\hat{\textbf{H}}_{img}]$, the integrated feature vectors, which is further activated with a softmax function to predict the likelihood over the four discourse labels.
%
%
For training, recall that in Table \ref{tab:stat}, we observe the severe label imbalance on our task. To deal with the issue, we adopt weighted cross-entropy loss, whose weights are set by the proportions of labels in training data.
\section{Experimental Setup}

\begin{table*}[t]
	 \centering
{\renewcommand{\arraystretch}{0.9}
\resizebox{1.9\columnwidth}{!}
{
	\begin{tabular}[b]{l c c c c c c}
		\toprule
		\textbf{Method}        & \textbf{Insertion} & \textbf{Concretization} &\textbf{Projection} & \textbf{Restatement} & \textbf{Extension} & \textbf{F1} \\
		\midrule
		\textbf{Baselines} \\
		\citet{2016CNN} 
		& 41.13   & 69.91  & 26.13  & 39.67 & 41.15 & 61.67  \\
		\citet{2016robust}    
		& 43.17   & 70.78  & 32.62  & 42.31 & 40.82 & 62.73  \\
		\citet{2017dual}    
		& 46.49   & 74.83  & 33.33  & 39.33 & 42.39 & 65.76   \\
		\hline
		\hline
		\textbf{Text+Image} \\
	    \textsc{ConcatFuse} & 52.86  & 81.62   & 34.78   & 39.19 &42.93 &71.09 \\
	    \textsc{Attention} & 54.30  & 82.64   & 33.71   & 39.23 & 39.41 & 71.48   \\

	    \textsc{Co-Attention} & 51.90  & 83.31   & 36.36   & 42.57 &40.59 &72.37  \\
	    \textsc{MultiheadAtt}& 53.69  & 84.33   & 36.96   & 42.11 &42.01 &73.33  \\
	    \hline
	    \textbf{Text+Caption} \\
	    \textsc{ConcatFuse} & 52.00  & 81.11   & 33.33   & 41.18 &43.02 &70.82  \\
	    \textsc{Attention} & 54.79  & 81.26   & 36.78   & 39.72 & 42.55 &70.97  \\
	    
	    \textsc{Co-Attention} & 53.73  & 82.13   & 37.20   &41.78 & 39.16 &71.38 \\

	    \textsc{MultiheadAtt}& 53.79  & 82.27   & 34.55   & 43.96 &43.46 & 72.08  \\
	    \hline
	    \textbf{Img+Text+Caption} \\
	    \textsc{ConcatFuse} & 52.48  & 82.41   & 32.97   & 43.01 &42.39 &71.88  \\
	    \textsc{Attention} & 53.24  & 83.01   & 34.95   & 43.45 & 43.65 & 72.58   \\
	    \textsc{Co-Attention} & 54.81  & 83.98   & 36.96   & 45.24 & 39.76 &73.15   \\

	    \textsc{MultiheadAtt}(\textit{full model}) & \textbf{57.75*}  & \textbf{84.88*}   & \textbf{37.36}   & \textbf{46.15*} &\textbf{44.19*} &\textbf{74.51*} \\
		\bottomrule	\end{tabular}}}
	\vspace{-0.5em}
	\caption{
	Comparison results of the baselines and our variants. Scores with * represent the significance tests of our full model over the baseline models with p-value$<$0.05.
	}
	\vspace{-1em}
	\label{tab:results}
\end{table*}
\paragraph{Model Settings.}
The length of tweet texts ($L$) and captions ($N$) are both capped at 20 by truncation. 
The batch size is set to 100, the learning rate to $5\times 10^{-5}$.
The head number of all multi-head attention layers are set to 6.
For image encoding, image feature map size $M$ is set to 14.
For text and comment encoding, the representations are extracted from the bottom 6-layers of the Bertweet model, which are further fine-tuned in training. 
In the setup, we randomly split 80\%, 10\% and 10\% for training, validation, and test. 
For evaluation, we report F1 scores in the prediction of each label and the weighted F1 to measure the overall results.


\paragraph{Baselines and Comparisons.}
We first consider two text-level discourse parsers proposed in \citet{2016CNN} and \citet{2016robust}, where we extend their text encoders into multimodal encoders to fit the image-text pairs. 
Then, we compare with a popular multimodal classifier~\cite{2017dual} %
that employs a dual attention network to fuse the visual and textual features.

Besides, we evaluate varying sets of feature combinations in our model
\emph{Test + Image}, \emph{Text + Caption}, and \emph{Text + Image + Caption} (the full set).
Recall that our framework employs multi-head attention to integrate features learned from different modalities. 
In experiments, we also test the performance of other modality fusion alternatives based on simple feature concatenation (\textsc{ConcatFuse}), the conventional attention mechanism (\textsc{Attention}), the co-attention mechanism (\textsc{Co-Attention}).


\section{Experimental Discussions}


This section first presents the main comparison results ( $\S$\ref{ssec:exp:main}).
Then, we discuss model sensitivity to varying modalities and text length in $\S$\ref{ssec:exp:length}.
Finally, $\S$\ref{ssec:exp:qualitative} presents a case study to provide more insights.

\subsection{Main Comparison Results}\label{ssec:exp:main}

Table \ref{tab:results} shows the main comparison results of various multimodal encoders. The following observations can be drawn.

First, all models do not exhibit good F1. This indicates that cross-modality discourse prediction is a challenging task. 
A good understanding for that cannot be gained by trivially adapting discourse parsers to the multimodal settings or applying
the existing vision-language encoders.
Second, results on the two entity-level discourse labels (i.e., insertion and concretization) are relatively better than scene-level, indicating that local objects are easier to be captured than global scenes.
Among all the labels, models perform the best in concretization, probably attributed to its richer data samples for feature learning (as shown in Table \ref{tab:stat}). And models obtains worst results in projection. The reasons might be that additional knowledge are needed for models to learn the implicit relation between the object and the entity.

Last, images, texts, and captions all contribute to building automatic discourse understanding. Joint modeling of the three modalities enables the corresponding models to outperform their text+image and text+caption counterparts.

\begin{figure}[t]
\centering
\includegraphics[width=0.8\columnwidth]{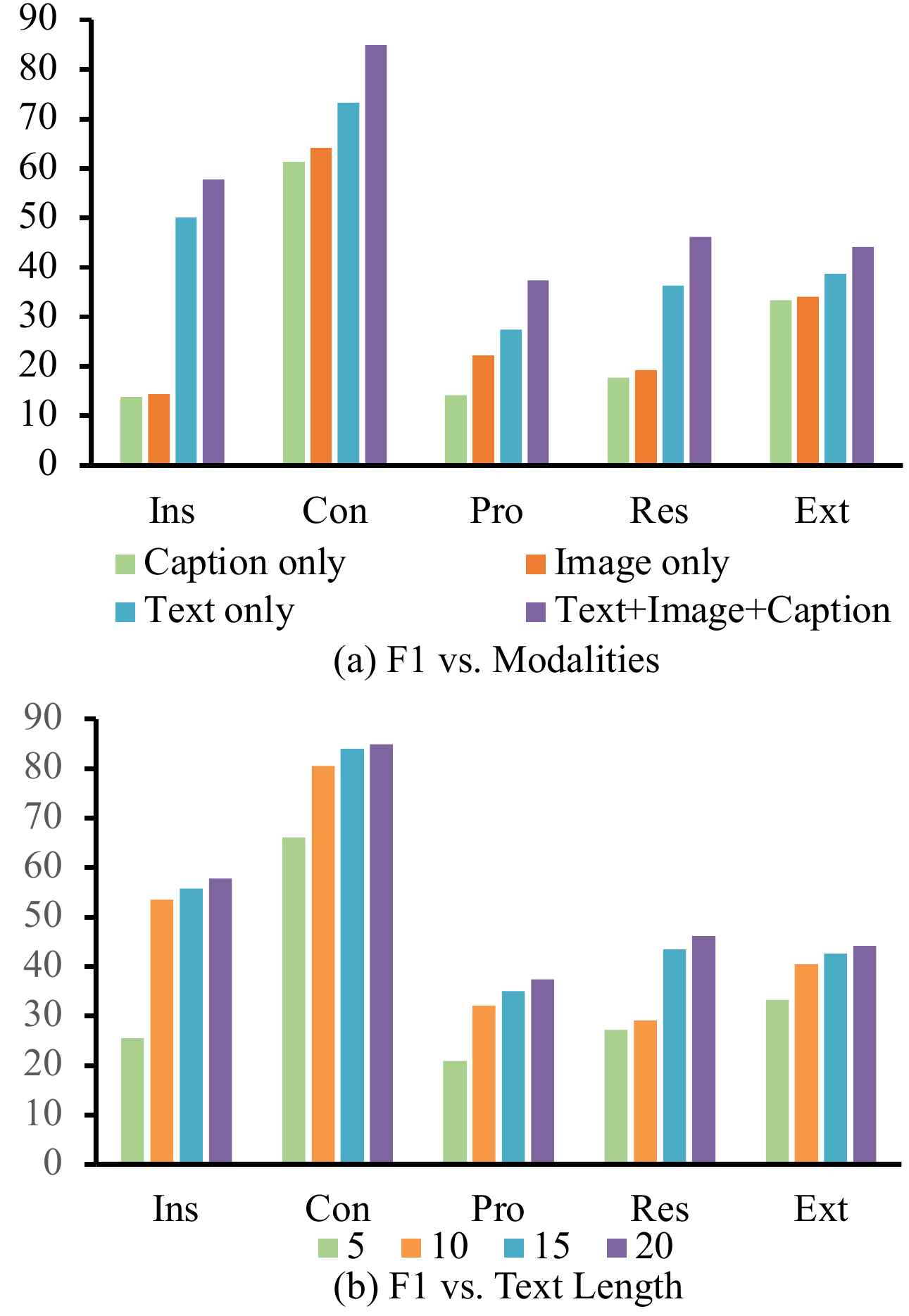}
\vspace{-0.5em}
\caption{
Full model performance compared with varying modality ablations in (a) and its results over varying text length (b). X-axis: insertion, concretization, projection, restatement, and extension; Y-axis: F1 scores. For each label, bars from left to right show the caption only, image only, text only ablations, and the full model in (a) and the tweet texts capped at 5, 10, 15, and 20 in (b). 
}
\vspace{-1em}
\label{fig:ablation}
\end{figure}

\subsection{Sensitivity to Modalities and Text Length}\label{ssec:exp:length}

\paragraph{Varying Modalities.}

To further examine the effects of varying modalities, we compare the F1 scores of our full model with its caption-only, image-only, and text-only ablations in Fig. \ref{fig:ablation}(a). 
It is seen that text modality contributes relatively more to discourse modeling observed from all labels, especially for insertion, where Name Entities are omitted and makes the text style easy to recognize.
Nevertheless, the joint effects of images, texts, and captions together present the best performance over all labels.

\paragraph{Varying Text Length.}
As discussed above, text features are crucial to predict cross-modality discourse. Here we further examine the effects of text length on model performance and the results of our full model are shown in Fig. \ref{fig:ablation}(b).
Better scores are observed for longer texts as richer contents can be captured.
This again demonstrates the essential signals provided by texts to infer cross-modality discourse.

\subsection{Qualitative Analysis}\label{ssec:exp:qualitative}


Discussions above mostly concern caption and text modalities.
Here we present a case study to probe into how the model reflects discourse indicators over vision signals.

\begin{figure}[t]
\centering
\includegraphics[width=1.0\columnwidth]{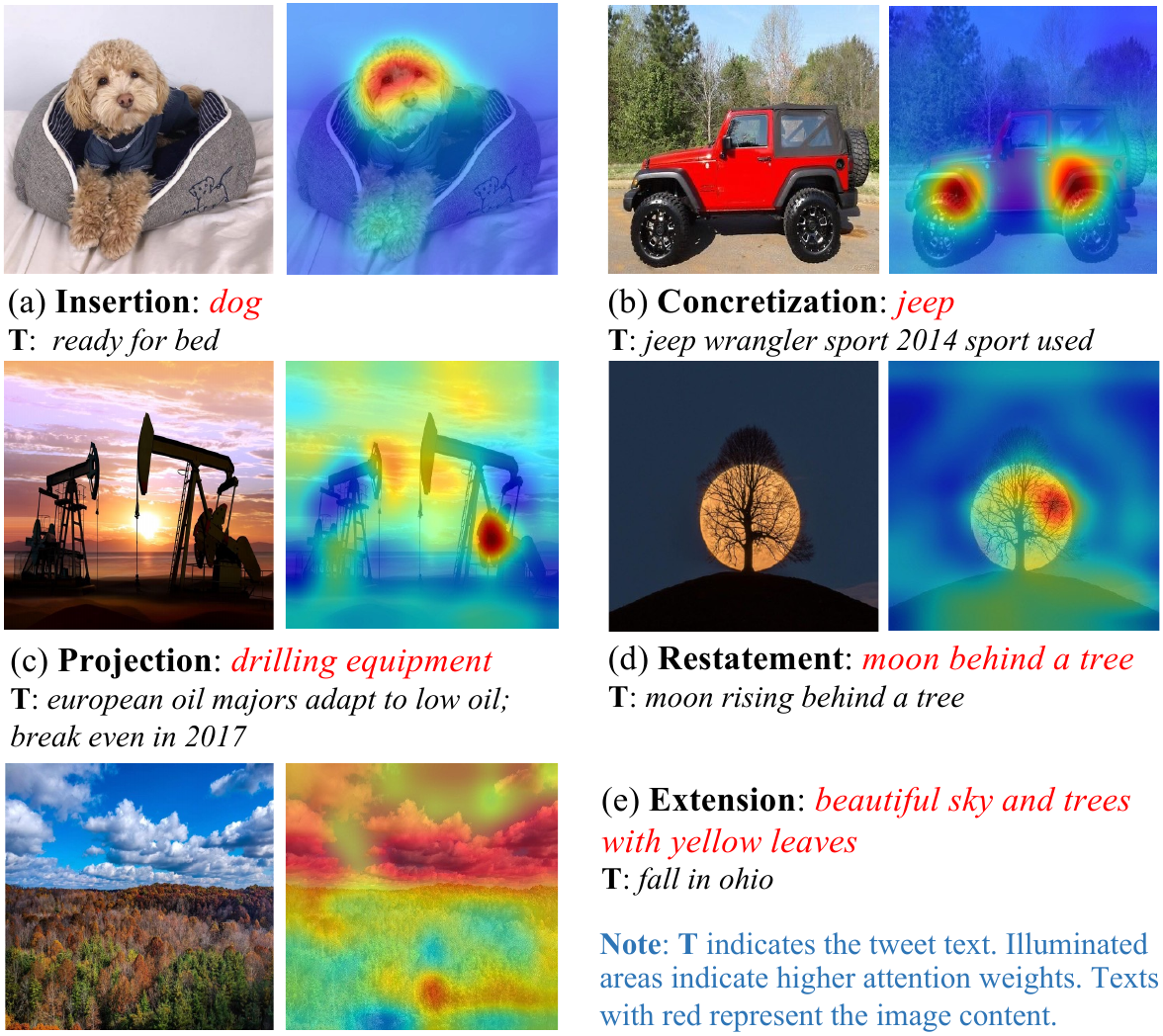}
\vspace{-1em}
\caption{
Visualization of multi-head attention heatmaps over sample images.
}
\vspace{-1.5em}
\label{fig:attention2}
\end{figure}

\paragraph{Case Study.}
Visual features are analyzed by the heatmap (in Fig. \ref{fig:attention2}) visualizing the text-aware attention weights over images (Eq. \ref{eq:head-attention}), which is  captured from image-text interactions.
As can be seen, attentions are able to highlight salient regions that signal the essential semantic links with the texts, e.g., the entities (dog and jeep) in (a) and (b).
It is also observed that the attention would vary in their focus in regions: for entity level discourse, it tends to concentrate on the some parts of a salient object (entity), while for scene level, attention also examines the background to capture the global view.

\section{Conclusion}
We have presented a novel task to learn cross-modality discourse that advances models to gain social media language understanding capability in multimedia contexts.
To handle the intricate image-text interactions, the visual semantics are first converted into text-formed subtitles and then compared with post texts to explore deep syntactic relations in linguistic space.
For empirical studies, we further contribute the first dataset presenting 16K human-annotated tweets with discourse labels for image-text pairs.
The main comparison results on our dataset have shown the effectiveness of multi-head attentions in exploring interactions among text, image, and caption modalities.
Further discussions demonstrate our potential to produce meaningful representations indicating implicit image-text structure.
These discourse features, conveying essential linguistic clues consistent with human senses, may largely benefit the future advances of automatic cross-modality understanding on social media.

\section*{Limitations}
Class imbalance is one of the main limitations of this work. As illustrated in Table \ref{tab:stat}, Concretization is the majority category which occupies 66.0\% of the dataset, while the minority categories, e.g. Projection and Insertion only account for 4.3\% and 5.2\% respectively. Although such uneven distribution reflects the real scenario of image-text relationships among tweets, future work should acquire a larger amount of minority categories for better interpretation of image-text relationships.

Cross-lingual and multi-platform studies should also be considered in later studies. It would be interesting and insightful to investigate the cross-modality discourse categories distribution among different languages. Are there any cultural traits that affect the use of image and text? Meanwhile, social media platforms can also exhibit preference for image and text usage. For example, will users on Instagram prefer to omit the Name Entities (Insertion category) than Twitter users? 

A more concrete model, e.g. vision-language Transformers, could also be employed to encode the text, caption, and image jointly. Current model runs efficiently on single NVIDIA RTX3080Ti GPU, while the training consumption of vision-language Transformers could be costly and requires larger dataset. Future studies could explore the trade-off between computation cost and classification performance.

\section*{Ethical Considerations}
We declare our dataset will cause no ethics problem. 
First, we follow the standard data acquisition process regularized by Twitter API. We downloaded the data for a purpose of academic research and is consistent with the Twitter terms of use. 
Then, we thoroughly navigated the data and ensured that no content will rise any ethics concerns, e.g. toxic languages, human face images, and censored images. 
Next, we perform the data anonymization to protect the user privacy. 
For the language use, we only keep the posts with English text. 
For the human annotations, we recruited the annotators as part-time research assistants with 16 USD/hour payment.
\section*{Acknowledgements}
This paper is substantially supported by NSFC Young Scientists Fund (No.62006203, 62106105), a grant from the Research Grants Council of the Hong Kong Special Administrative Region, China (Project No. PolyU/25200821), PolyU internal funds (1-BE2W, 4-ZZKM, and 1-ZVRH), and CCF-Baidu Open Fund (No. 2021PP15002000).

\bibliographystyle{acl_natbib}
\bibliography{anthology,custom}
\end{document}